\documentclass[aps,pra,twocolumn]{revtex4}
\usepackage{graphicx}
\usepackage{dcolumn}
\usepackage{bm}
\usepackage[]{hyperref}
\usepackage{color} 

\begin{document}


\title{Low-light-level four-wave mixing by quantum interference}
\author{Chang-Kai Chiu,$^{1}$ Yi-Hsin Chen,$^{2}$ Yen-Chun Chen,$^{1}$ Ite A. Yu,$^{2}$ Ying-Cheng Chen,$^{3}$ and Yong-Fan Chen$^{1,}$}

\email{yfchen@mail.ncku.edu.tw}

\affiliation{$^1$Department of Physics, National Cheng Kung University, Tainan 70101, Taiwan \\
$^2$Department of Physics and Frontier Research Center on Fundamental and Applied Sciences of Matters, National Tsing Hua
University, Hsinchu 30013, Taiwan \\
$^3$Institute of Atomic and Molecular Sciences, Academia Sinica, Taipei 10617, Taiwan}



\begin{abstract}
We observed electromagnetically-induced-transparency-based four-wave mixing (FWM) in the pulsed regime at low light levels. The FWM conversion
efficiency of $3.8(9)\%$ was observed in a four-level system of cold $^{87}{\rm Rb}$ atoms using a driving laser pulse with a peak intensity of
$\approx$ 80 ${\rm \mu W/cm^2}$, corresponding to an energy of $\approx$ 60 photons per atomic cross section. Comparison between the
experimental data and the theoretical predictions proposed by Harris and Hau [Phys. Rev. Lett. 82, 4611 (1999)] showed good agreement.
Additionally, a high conversion efficiency of $46(2)\%$ was demonstrated when applying this scheme using a driving laser intensity of $\approx$
1.8 ${\rm mW/cm^2}$. According to our theoretical predictions, this FWM scheme can achieve a conversion efficiency of nearly $100\%$ when using
a dense medium with an optical depth of 500.
\end{abstract}



\maketitle

\newcommand{\FigOne}{
    \begin{figure}[t] 
    \includegraphics[width=9.00cm]{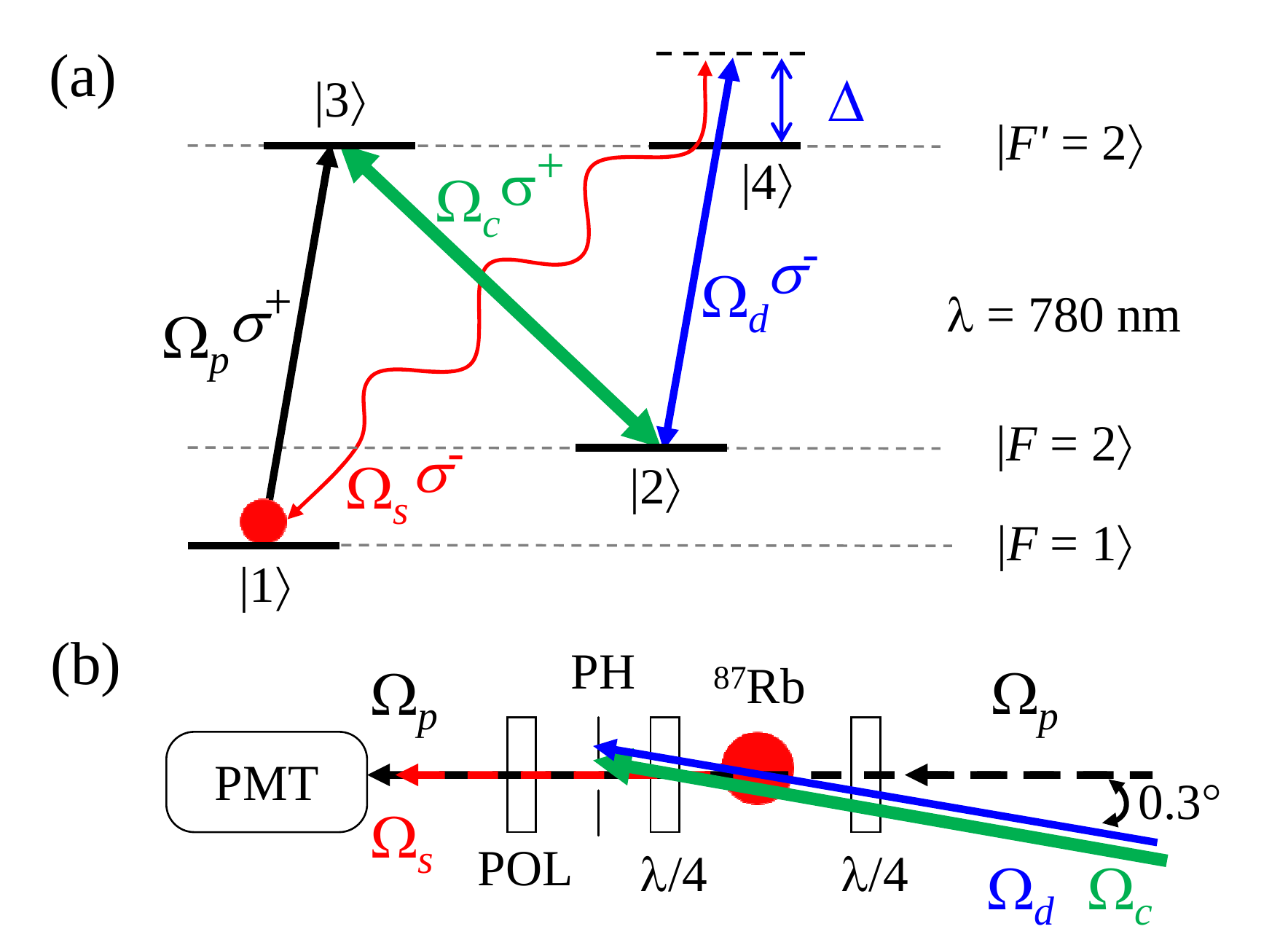}
    \caption{(color online).
Energy level scheme and experimental apparatus. (a) Energy levels of $^{87}{\rm Rb}$ $D_2$-line transition for the EIT-based FWM experiment.
Driving detuning, $\Delta$, is defined as $\omega_d - \omega_{24}$, where $\omega_d$ and $\omega_{24}$ are the frequencies of the driving field
and the $|2\rangle \leftrightarrow |4\rangle$ transition, respectively. (b) Schematic diagram of the experimental setup. $\lambda$/4,
quarter-wave plate; POL, polarizer; PMT, photomultiplier tube; and PH, 200-$\rm\mu$m pinhole for blocking the scattering light of the coupling
and driving fields.}
    \label{fig:setup}
    \end{figure}
}
\newcommand{\FigTwo}{
    \begin{figure}[t] 
    \includegraphics[width=9.0cm]{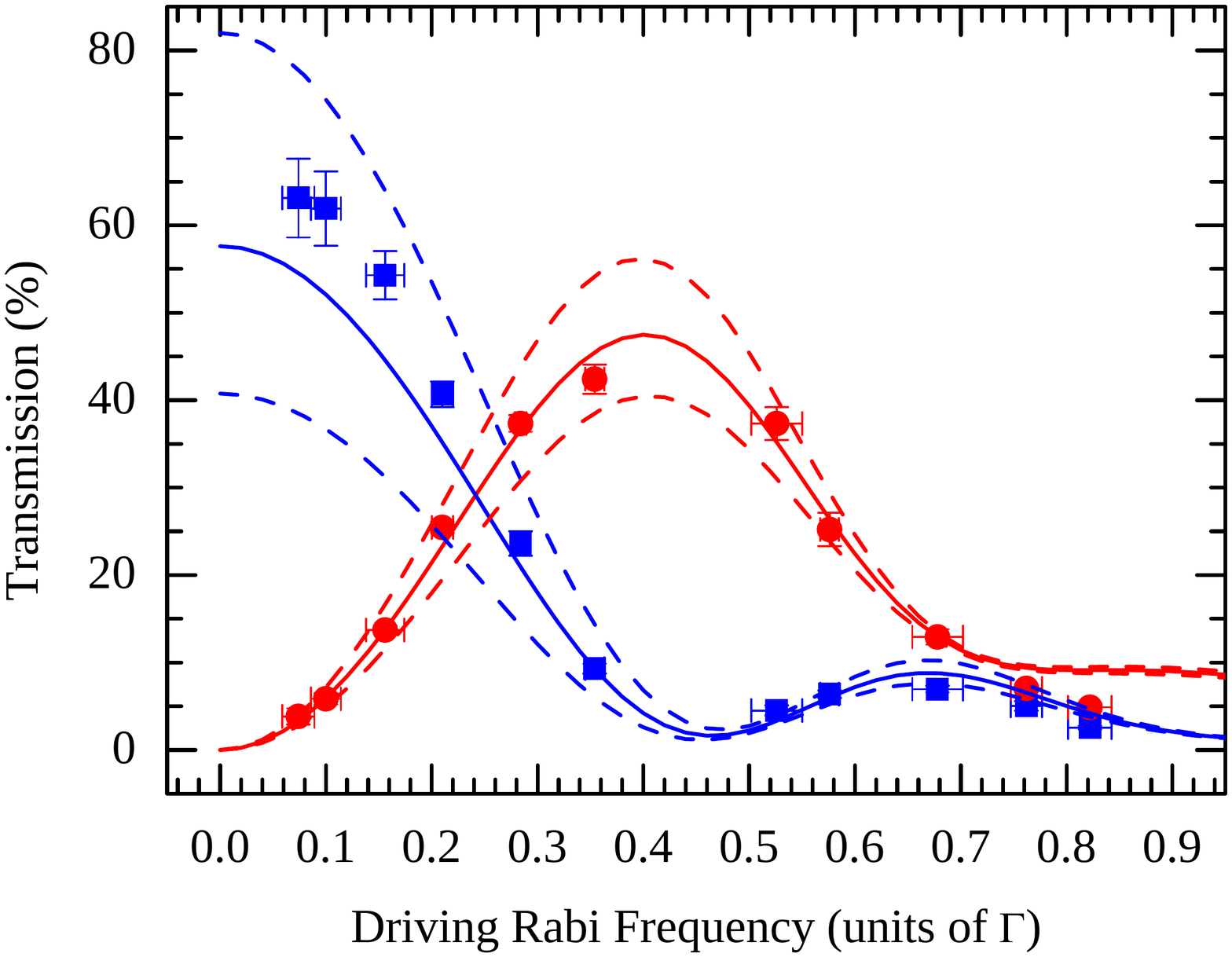}
    \caption{(color online).
Four-wave mixing transmission of the probe and signal pulses as a function of the driving Rabi frequency ($\Omega_{d}$). The blue squares and
the red circles show the transmission of the probe and the signal pulses, respectively. The solid lines show the theoretical curves with the
parameters $\Delta$ = $13\Gamma$, $\alpha$ = 42, $\Omega_c=0.32\Gamma$ and, $\gamma_{21}=9\times 10^{-4}\Gamma$. Consider all data in this
measurement, $\gamma_{21}$ has a standard deviation of $7\times 10^{-4}\Gamma$; hence we also plot the theoretical curves (dashed lines) with
the parameters $\gamma_{21}=1.6\times 10^{-3}\Gamma$ (lower dashed lines) and $\gamma_{21}=2\times 10^{-4}\Gamma$ (upper dashed lines). The
error bars represent $\pm$ 1 standard deviation based on measurement statistics.}
    \label{fig:Rabi}
    \end{figure}
}
\newcommand{\FigThree}{
    \begin{figure}[t] 
    \includegraphics[width=9.0cm]{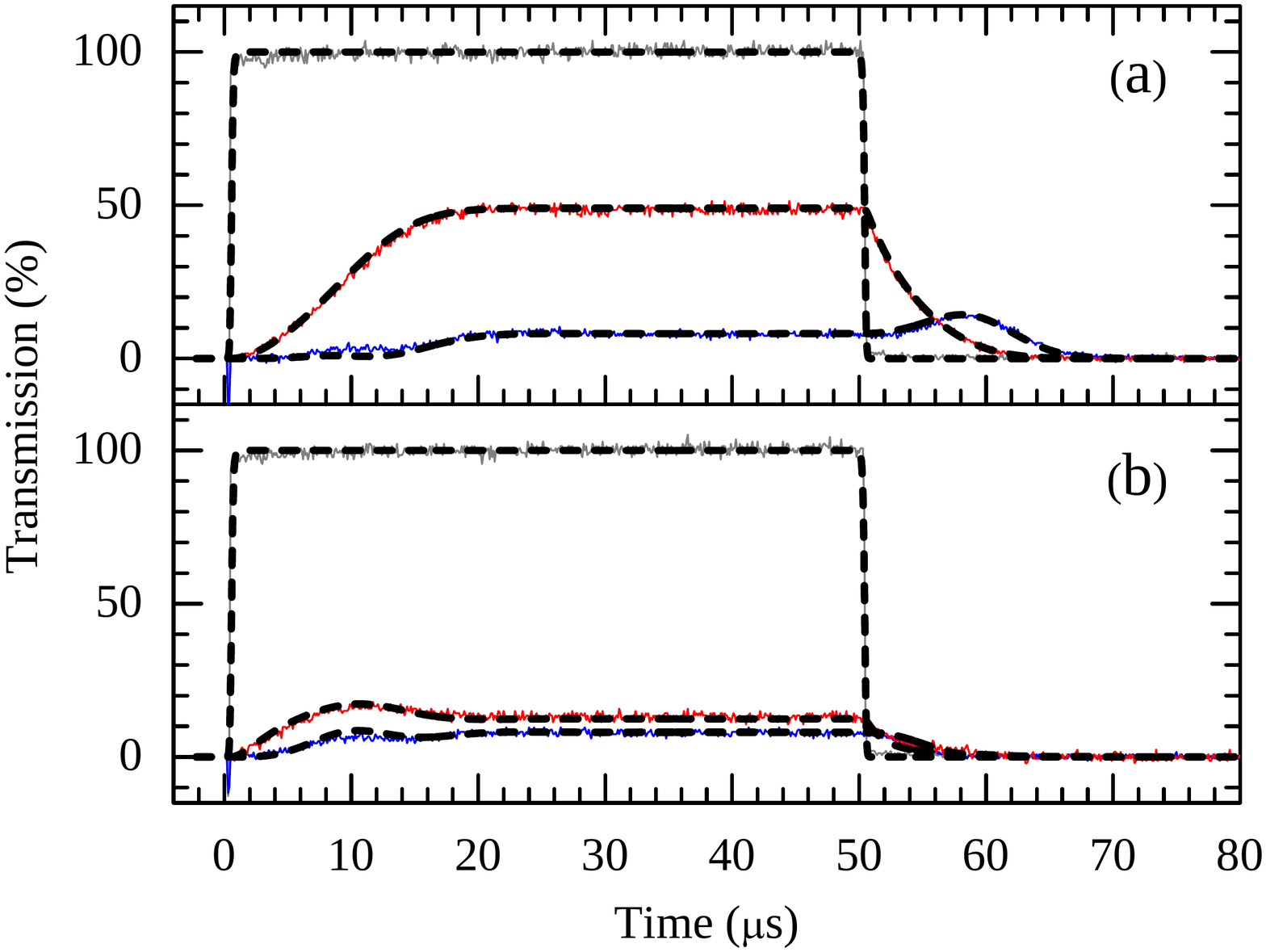}
    \caption{(color online).
Probe and signal pulses propagating through the EIT-based FWM medium. The gray (upper) lines are the incident probe pulses. The red (middle) and
blue (lower) lines are the generated signal and the transmitted probe pulses, respectively. The dashed lines plot the theoretical curves with
the parameters (a) $\Delta$ = $13\Gamma$, $\alpha=40$, $\Omega_c=0.32\Gamma$, $\Omega_d=0.37\Gamma$, and $\gamma_{21}=1\times 10^{-3}\Gamma$.
(b) $\Delta$ = $13\Gamma$, $\alpha= 42$, $\Omega_c=0.32\Gamma$, $\Omega_d=0.67\Gamma$, and $\gamma_{21}=1.5\times 10^{-3}\Gamma$. The FWM
conversion efficiencies in (a) and (b) were 42$\%$ and 13$\%$, respectively.}
    \label{fig:Time}
    \end{figure}
}
\newcommand{\FigFour}{
    \begin{figure}[t] 
    \includegraphics[width=9.0cm]{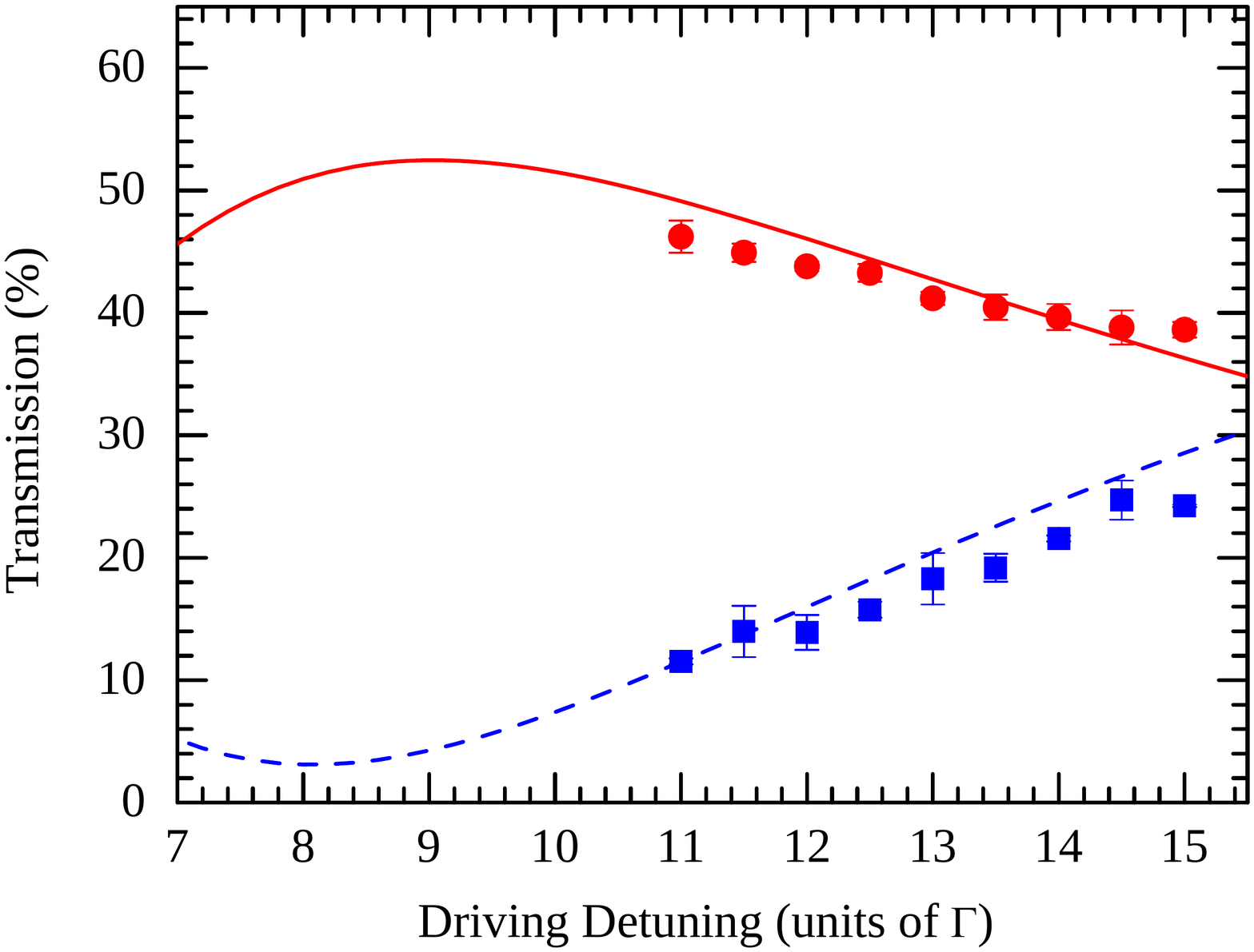}
    \caption{(color online).
Four-wave mixing conversion efficiency as a function of the driving detuning ($\Delta$). The blue squares and red circles show the transmission
of the probe and signal pulses, respectively. The dashed blue (probe) and solid red (signal) lines are the theoretical curves with the
parameters $\alpha= 36$, $\Omega_c=0.32\Gamma$, $\Omega_d=0.35\Gamma$, and $\gamma_{21}=6\times 10^{-4}\Gamma$. The error bars represent $\pm$ 1
standard deviation based on measurement statistics.}
    \label{fig:detuning}
    \end{figure}
}

\newcommand{\FigFive}{
    \begin{figure}[t] 
    \includegraphics[width=9.0cm]{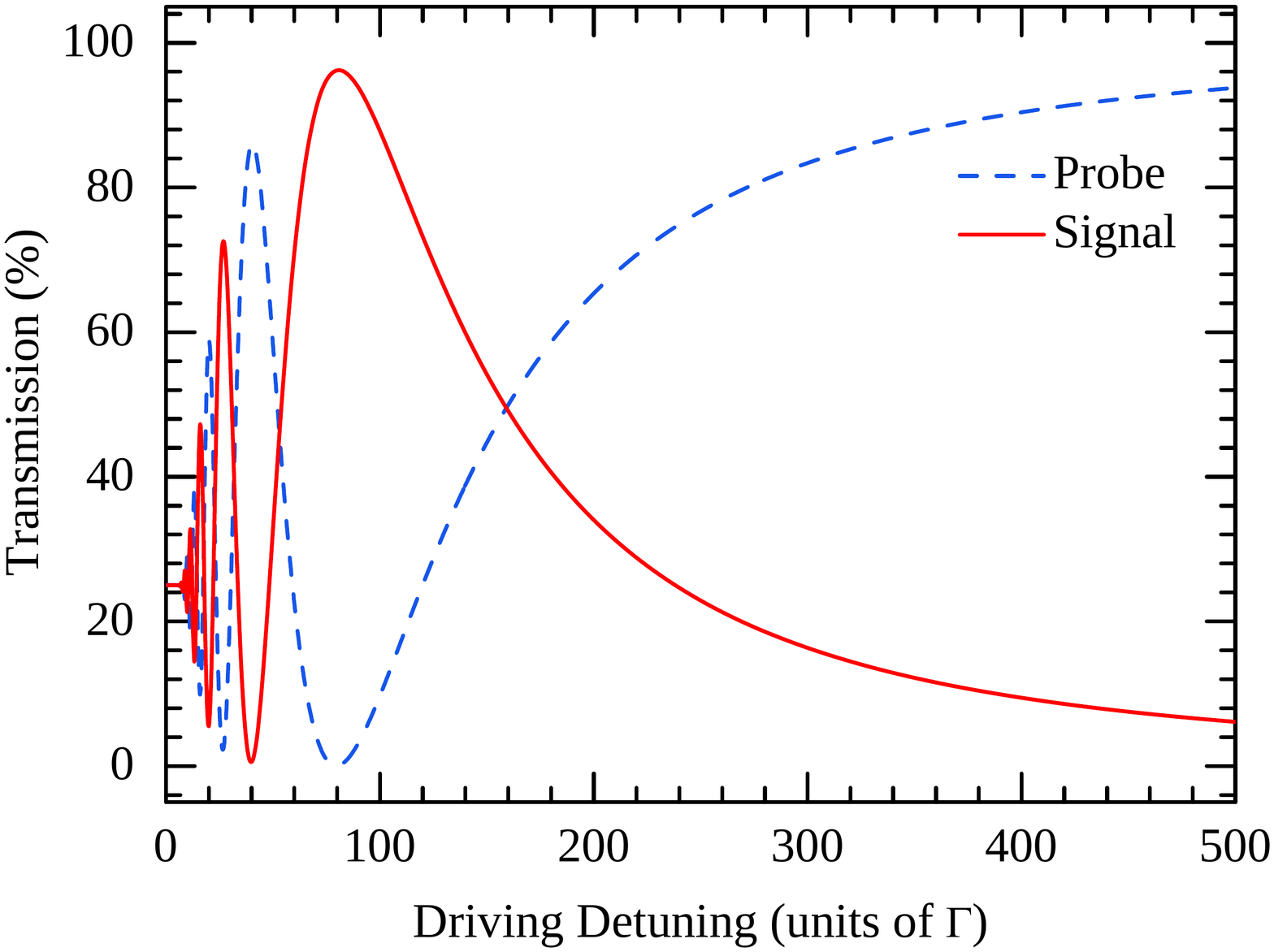}
    \caption{(color online).
Nearly $100\%$ FWM conversion efficiency with a great optical depth. The dashed blue (probe) and solid red (signal) lines are the theoretical
curves plotted according to Eqs.~(\ref{Eq.probe}) and (\ref{Eq.signal}) with the parameters $\alpha= 500$ and $\Omega_c = \Omega_d
=0.32\Gamma$.}
    \label{fig:efficiency}
    \end{figure}
}


\section{Introduction}

Suppressing linear absorption and enhancing nonlinear interaction by using electromagnetically induced transparency (EIT) has been suggested to
be a potential technique for all-optical control at low light levels~\cite{Harris99,EIT05,KangXPM,BrajePS,ChenPS,Chen06,Chen10}. To date,
efficient all-optical switching and cross-phase modulation based on EIT have been realized at the level of a few hundred
photons~\cite{Bajcsy09,YFC11}. Recently, optical switching at the single-photon level based on cavity EIT was demonstrated~\cite{Chen13}.
Another crucial nonlinear effect based on EIT is four-wave mixing (FWM) which can be used to efficiently change the frequency of a photon. An
individual photon with a tunable frequency can carry information among various types of devices, and may have potential applications in quantum
networks~\cite{Kimble08,kuzmich10}. However, to be useful in quantum control and communications, FWM must be highly efficient and not
introducing additional noise photons~\cite{Raymer10,Zeilinger11,Raymer12}.

Here, we report an experimental observation of EIT-based FWM in the pulsed regime at low light levels. An FWM conversion efficiency of $46(2)\%$
was demonstrated using a driving laser pulse with a peak intensity of $\approx$ 1.8 ${\rm mW/cm^2}$ in a four-level system of cold $^{87}{\rm
Rb}$ atoms. According to our theoretical predictions, this FWM scheme can achieve a conversion efficiency of nearly $100\%$ when using a dense
medium with a great optical depth at low light levels. In addition, we observed a conversion efficiency of $3.8(9)\%$ when using a driving laser
intensity of $\approx$ 80 ${\rm \mu W/cm^2}$, corresponding to an energy of $\approx$ 60 photons per atomic cross section. Comparison between
the experimental data and the theoretical predictions proposed by Harris and Hau~\cite{Harris99} showed good agreement.

Before describing this study, we review previous experiments. Zhang \emph{et al.} first observed sum-frequency generation by using EIT in atomic
hydrogen~\cite{Zhang93}. Jain \emph{et al.} demonstrated an efficient frequency conversion by using EIT in atomic Pb vapor~\cite{Jain96}.
However, the laser intensities used in the two experiments were in the order of ${\rm MW/cm^2}$. Merriam \emph{et al.} demonstrated an efficient
gas-phase generation of ultraviolet radiation by using a double-$\Lambda$ system in the pulsed regime~\cite{Merriam99,Merriam00}. Braje \emph{et
al.} reported first experimental demonstration of frequency conversion by using EIT in clod atoms~\cite{Braje04}. Kang \textit{et al.}
demonstrated an FWM conversion efficiency of $10\%$ by using EIT in the pulsed regime~\cite{Kang04}, and, recently, Wang \textit{et al.}
improved the efficiency to 73$\%$ in hot Rb atoms by using continuous-wave lasers with intensities greater than ${\rm W/cm^2}$~\cite{Wang10}.

\section{Theoretical Model} \label{Sec:thy}

In the present study, we consider a four-level FWM system based on EIT, as shown in Fig.~\ref{fig:setup}(a). The behavior of the probe and the
generated signal pulses propagating in the medium were theoretically analyzed by solving the Maxwell-Schr\"{o}dinger equations
\begin{eqnarray}
\frac{\partial\Omega_{p}}{\partial z} + \frac{1}{c}\frac{\partial\Omega_{p}}{\partial t} &= i \frac{\alpha_{p}\gamma_{31}}{2L} \rho_{31},\label{Eq.slowP}\\
\frac{\partial\Omega_{s}}{\partial z} + \frac{1}{c}\frac{\partial\Omega_{s}}{\partial t} &= i \frac{\alpha_{s}\gamma_{41}}{2L}
\rho_{41},\label{Eq.slowS}
\end{eqnarray}
where $\Omega_p$ and $\Omega_s$ are the Rabi frequencies of the probe and signal pulses, respectively; $\rho_{ij}$ is the slowly varying
amplitude of the coherence between states $|i\rangle$ and $|j\rangle$; $\gamma_{31}$ and $\gamma_{41}$ are the total coherence decay rates from
the excited states $|3\rangle$ and $|4\rangle$, respectively; and $\alpha_{p} = n \sigma_{13}L$ ($\alpha_{s} = n\sigma_{14}L$) represents the
optical depth of the probe (signal) transition, where $n$ is the number density of the atoms, $\sigma_{13}$($\sigma_{14}$) is the atomic
absorption cross section of the probe (signal) transition, and $L$ is the optical path length of the medium. The optical depths of the probe and
signal transitions in this experiment were the same ($\alpha_{p}=\alpha_{s}=\alpha$) because $\sigma_{13}$ is equal to $\sigma_{14}$ when three
degenerate Zeeman sublevels are considered.

When the probe and signal fields are very weak (i.e., $\rho_{11} \simeq 1$), the optical Bloch equations of the slowly varying amplitudes of the
density-matrix elements are given by
\begin{eqnarray}
\frac{d}{dt}\rho_{41} = \frac{i}{2}\Omega_{s} + \frac{i}{2}\Omega_{d}\rho_{21} + \left(i\Delta -
\frac{\gamma_{41}}{2}\right)\rho_{41},\\\label{Eq.p41} \frac{d}{dt}\rho_{31} = \frac{i}{2}\Omega_{p} + \frac{i}{2}\Omega_{c}\rho_{21} -
\frac{\gamma_{31}}{2}\rho_{31},\\\label{Eq.p31} \frac{d}{dt}\rho_{21} = \frac{i}{2}\Omega^{\ast}_{c}\rho_{31} +
\frac{i}{2}\Omega^{\ast}_{d}\rho_{41} - \frac{\gamma_{21}}{2}\rho_{21},\label{Eq.p21}
\end{eqnarray}
where $\Delta$ denotes the detuning of the driving transition [see Fig.~\ref{fig:setup}(a)]. The parameter $\gamma_{21}$ represents the
dephasing rate of ground states $|1\rangle$ and $|2\rangle$. Each parameter in the theoretical model was individually determined in additional
experiments as follows: $\Omega_c$ was measured based on the separation of the two absorption peaks in the EIT spectrum~\cite{Harris97};
$\alpha$ was measured based on the delay time of the slow light pulse~\cite{HauSlowLt}; $\gamma_{31}$ and $\gamma_{41}$, which were contributed
mainly by the spontaneous decay rate and laser linewidth, were both $1.25(2)\Gamma$ according to the spectral width of the one-photon
absorption~\cite{Lo10}. The spontaneous decay rate of the excited states was $\Gamma = 2\pi \times 6$ MHz; and $\Omega_d$ was derived by
numerically fitting the transmitted profiles of both the probe and signal pulses propagating through the FWM medium, and we modified
$\gamma_{21}$ to obtain the best fit of the amplitudes of the transmitted pulses.

\FigOne

When $\gamma_{21} = 0$ and $\gamma_{31}$ = $\gamma_{41}$, the steady-state solutions for the probe and signal fields are obtained by solving
Eqs.~(\ref{Eq.slowP})-(\ref{Eq.p21}) as follows:
\begin{eqnarray}
\Omega_p(\alpha) = \frac{\Omega_p(0)}{\Omega^{2}} {\left[\Omega_c^{2} + \Omega_d^{2}e^{-i\frac{\alpha}{2\xi}}\right]},\label{Eq.probe}\\
\Omega_s(\alpha) = \frac{\Omega_p(0)}{\Omega^{2}} {\left[\Omega_c\Omega_d - \Omega_c\Omega_de^{-i\frac{\alpha}{2\xi}}\right]},\label{Eq.signal}
\end{eqnarray}
where $\Omega^{2} \equiv \Omega_c^{2}+\Omega_d^{2}$ and $\xi=i+2\frac{\Omega_c^{2}\Delta}{\Omega^{2}\gamma_{31}}$. The steady-state conversion
efficiency is defined as the intensity ratio of the generated signal field to the incident probe field. According to Eqs.~(\ref{Eq.probe}) and
(\ref{Eq.signal}), for a given $\Omega_c$ and $\Omega_d$, the conversion efficiency under the condition of three-photon resonance ($\Delta=0$)
saturates at a certain value for a sufficiently great optical depth, and the optimal value of 25$\%$ is achieved when $\Omega_c=\Omega_d$. The
EIT-based FWM scheme also allows a weak driving field to yield a conversion efficiency higher than 25$\%$ when $\Delta \neq 0$~\cite{Deng02}.
Moreover, using a great optical depth ($\approx$ 500) with a moderate $\Delta$, $\Omega_c$, and $\Omega_d$ enable a conversion efficiency of
nearly $100\%$ to be achieved when using this FWM scheme, which is discussed in the section of results and discussion.




\section{Experimental Details} \label{Sec:exp}

The experiment was conducted in a magneto-optical trap (MOT) of $^{87}$Rb. To increase the optical depth, we designed a dark spontaneous-force
optical trap (SPOT)~\cite{Ketterle93}. Details on the experimental setup of the dark SPOT are provided in our previous study~\cite{YFC13}. In
the present study, the relevant energy levels of $^{87}$Rb atoms are shown in Fig.~\ref{fig:setup}(a). All populations were prepared in ground
state $|1\rangle$ ($|5S_{1/2}, F=1\rangle$). The probe ($\Omega_p$) and coupling ($\Omega_c$) fields with $\sigma^+$ polarization drove the
transitions of states $|1\rangle\rightarrow |3\rangle$ and $|2\rangle\rightarrow |3\rangle$, respectively, to form a $\Lambda$-type EIT. The
driving field $\Omega_d$ with $\sigma^-$ polarization drove the transition of states $|2\rangle\rightarrow |4\rangle$, and had a frequency
detuning $(\Delta)$. State $|2\rangle$ was a ground state of $|5S_{1/2}, F=2\rangle$, and states $|3\rangle$ and $|4\rangle$ were degenerated
excited states ($|5P_{3/2}, F'=2\rangle$). We note that the Zeeman splitting of the hyperfine sublevels in this experiment was reduced to a few
kHz or less because the background magnetic field was compensated to below 1 mG by using three pairs of magnetic coils.

The coupling and driving fields were produced by one diode laser which was injection locked by an external cavity diode laser (ECDL). One beam
from the ECDL was sent through a 6.8-GHz electro optic modulator before seeding the probe field. The peak power of the probe pulse was set to
$\approx$ 1 nW which is much smaller than that of the coupling pulse ($\approx$ 60 $\rm \mu W$) throughout the experiment. Therefore, most of
the populations would stay in the ground state $|1\rangle$ even though at nearly $100\%$ FWM efficiency. The probe and coupling beams exhibited
waist diameters of approximately 0.2 and 2.2 mm, respectively. The beam diameter of the driving field was the same as that of the coupling beam
in the experiment.

\FigTwo
\FigThree

The propagating direction of the probe beam was denoted as $\theta=0^\circ$. The coupling and driving beams were coupled by a single-mode fiber
and were then sent along the direction of $\theta \approx 0.3^\circ$. Because of the FWM effect, the signal field $\Omega_s$, which is generated
when the phase matching condition is satisfied, propagates along with the direction of the probe beam [see Fig.~\ref{fig:setup}(b)]. The probe
and signal fields were detected using a photomultiplier tube, but were distinguished using a polarizer. Most of the scattering light from the
coupling and driving beams was blocked by a 200-$\rm\mu$m pinhole.

In the timing sequence, we denoted the time of firing the probe pulse as $t = 0$. The magnetic field of the MOT was first turned off at $t = -2$
ms. We then switched off the repumping laser of the MOT at $t = -210~\mu$s and immediately turned on the coupling field. The trapping laser of
the MOT was switched off at $t=-2~\mu$s before sending the probe and driving pulses simultaneously. This process was used to ensure that the
entire population was optically pumped to the ground state $|1\rangle$. The timing and detuning ($\Delta$) of all light fields were controlled
using individual acousto optic modulators. The repetition rate for each measurement was 100 Hz. Throughout the experiment, the duration of the
probe square pulse, $T_p$, was set to $50~\mu$s to satisfy the condition that the spectral width of the probe pulse must be much smaller than
the width of EIT transparency window; moreover, the driving square pulse with a duration of $T_{24} = 70~\mu$s was used to cover the entire
slow-light probe pulse.

\section{Results and Discussion} \label{Sec:result}

The conversion efficiency of EIT-based FWM is a function of $\Omega_c$, $\Omega_d$, $\alpha$, and $\Delta$, as shown in Eqs.~(\ref{Eq.probe})
and (\ref{Eq.signal}). We first measured the conversion efficiency as a function of $\Omega_d$ when $\Delta$ = 13$\Gamma$, $\alpha$ = 42, and
$\Omega_c$ = 0.32$\Gamma$. The blue squares and red circles in Fig.~\ref{fig:Rabi} represent the experimental data of the transmitted probe and
signal pulses, respectively, and the solid lines are the theoretical curves estimated by numerically solving
Eqs.~(\ref{Eq.slowP})-~(\ref{Eq.p21}) where the dephasing rate $\gamma_{21}$ was set to $9\times 10^{-4}\Gamma$. Consider all data in
Fig.~\ref{fig:Rabi}, $\gamma_{21}$ has a standard deviation of $7\times 10^{-4}\Gamma$. The large deviation of the dephasing rate is probably
attributed to an increase of the strength of the driving field, which increased the dephasing rate, because the driving field also drives the
transition of state $|F=2\rangle$ to state $|F'=3\rangle$, resulting in a far-off resonance photon-switching effect \cite{memory13}. This
unwanted effect can be avoided by using the $D_1$-line transitions of alkali atoms. The dashed lines (numerical curves) are guidelines for the
experimental data with $\gamma_{21}=1.6\times 10^{-3}\Gamma$ (lower dashed lines) and $\gamma_{21}=2\times 10^{-4}\Gamma$ (upper dashed lines).
For a weaker (stronger) driving field, the transmission of these two pulses approaches the upper (lower) dashed line. The experimental data also
show that signal transmission exhibited a maximum value when $\Omega_d \approx \Omega_c$. An FWM conversion efficiency of $42(2)\%$ was obtained
when $\Omega_d=0.35\Gamma$, corresponding to a driving peak intensity of $\approx$ 1.8 ${\rm mW/cm^2}$.


We also observed an FWM conversion efficiency of $3.8(9)\%$ (the leftmost data in Fig.~\ref{fig:Rabi}) when using a driving pulse of $\Omega_d$
= 0.074$\Gamma$. The corresponding peak intensity, calculated as $I_d$ = $\left(2\Omega_d^2 I_0/\Gamma^2 a_{ij}^2\right)$, was $\approx$ 80
${\rm \mu W/cm^2}$, where $I_0$ $\approx$ 1.63 ${\rm mW/cm^2}$ is the saturation intensity and $a_{ij}^2=2/9$ is the average value of the square
of the Clebsch-Gordan coefficients determined by considering three degenerated Zeeman sublevels. The duration of the driving pulse, $T_{24}$,
was set to $70~\mu$s in this experiment. We estimated that the driving pulse carries an energy of $\approx$ 60 photons per atomic cross section
$(3\lambda^{2}/2\pi)$.

\FigFour

We subsequently compared the experimental data and the theoretical predictions proposed by Harris and Hau~\cite{Harris99}. According to their
theoretical analysis, the EIT-based FWM conversion efficiency, $\zeta$, in the pulsed regime is given by
\begin{eqnarray}
\zeta = N_{d} \frac{\sigma_{24}}{A}\Phi\left(\eta, r\right),\label{Eq.zeta}
\end{eqnarray}
where $N_d$ is the number of driving photons; $\sigma_{24}$ is the atomic cross section of the $|2\rangle \leftrightarrow |4\rangle$ transition;
and $A$ is the spot area of the signal beam. The term $\zeta$ is determined according to the dimensionless parameter, $\Phi\left(\eta,
r\right)$, where $\eta = T_{d}/T_{p}$ is the ratio of the EIT delay time to the probe pulse duration, and $r =
\frac{\gamma_{31}^{2}}{8\Delta^{2}+2\gamma_{31}^{2}}\alpha$ is equal to the loss of light propagating in the medium. For the experimental
parameters in Fig.~\ref{fig:Rabi}, we obtained $\eta = 0.27$ and $r =0 .048$; hence the analytic solution from~\cite{Harris99} shows
$\Phi=4.5\times 10^{-3}$. When considering a driving pulse containing 60 photons is tightly focused to a spot area of an atomic cross section in
the FWM experiment, we calculated a conversion efficiency of $6.0\%$ by using the Eq.~(\ref{Eq.zeta}). The difference between the experimental
data ($3.8\%$) and the theoretical value ($6.0\%$) was due to the square profiles of the probe and driving pulses and the non-negligible
dephasing rate ($\gamma_{21} \neq 0$) in the experiment.

Figures~\ref{fig:Time}(a) and \ref{fig:Time}(b) show typical data of the probe (middle) and signal (lower) pulses propagating through the FWM
medium with the driving Rabi frequencies of $0.37\Gamma$ and $0.67\Gamma$, respectively. The gray (upper) lines are the incident probe pulses.
The group velocity differed between the transmitted probe and the signal pulses because $\Delta \neq 0$. The FWM conversion efficiencies were
$42\%$ and $13\%$ in Figs.~\ref{fig:Time}(a) and~\ref{fig:Time}(b), respectively.


\FigFive

We further optimized the FWM conversion efficiency by modifying the driving detuning $\Delta$. We measured the conversion efficiency as a
function of $\Delta$ when $\alpha= 36$, $\Omega_c = 0.32\Gamma$, and $\Omega_d = 0.35\Gamma$, as shown in Fig.~\ref{fig:detuning}. The
experimental data show that the conversion efficiency increased as $\Delta$ decreased. A high conversion efficiency of 46(2)$\%$ was obtained
when $\Delta$ = 11$\Gamma$. According to the theoretical predictions shown in the solid line, a greater conversion efficiency of $52\%$ can be
achieved if the driving detuning is reduced from 11$\Gamma$ to 9$\Gamma$.

Figure~\ref{fig:efficiency} shows that an FWM conversion efficiency of nearly $100\%$ can be achieved when the optical depth is sufficiently
great~\cite{Peters13}. The dashed blue (probe) and solid red (signal) lines are the theoretical curves plotted according to
Eqs.~(\ref{Eq.probe}) and (\ref{Eq.signal}) with the parameters $\alpha= 500$ and $\Omega_c = \Omega_d =0.32\Gamma$. A nearly $100\%$ FWM
efficiency was obtained at low light levels when the driving detuning, $\Delta$, was set to approximately 80$\Gamma$. An efficient frequency
conversion of single photons can protect quantum information carried by photons and may have potential applications in quantum control and
communications~\cite{Raymer12}.


\section{CONCLUSION} \label{Sec:conclusion}

We experimentally demonstrated EIT-based FWM in the pulsed regime at low light levels. A conversion efficiency of $3.8(9)\%$ was observed when
using an ultraweak driving pulse with an energy of $\approx$ 60 photons per atomic cross section. We also demonstrated a high conversion
efficiency of $46(2)\%$ when using this scheme. According to our theoretical predictions, this FWM scheme can achieve a conversion efficiency of
nearly $100\%$ when using a dense medium with an optical depth of 500. Additionally, the experimental data confirmed the analytic solution
derived by the authors in~\cite{Harris99}.



\section*{ACKNOWLEDGEMENTS}

The authors thank Hao-Chung Chen for helpful discussions and Jun-Xian Chen for experimental assistance. This work was supported by the National
Science Council of Taiwan under Grant No. 101-2112-M-006-004-MY3.







\end{document}